\documentclass[aps,twocolumn,prb,reprint,amsmath,amssymb,showpacs,superscriptaddress]{revtex4-1}

\usepackage{graphicx}
\usepackage{dcolumn}
\usepackage{bm}
\usepackage{amsmath}
\usepackage{color}
\usepackage{mathrsfs}
\usepackage{float}
\usepackage{dsfont}
\usepackage{txfonts}
\usepackage{wasysym}
\usepackage{ifsym}

\makeatletter

\newcommand{\Rmnum}[1]{\expandafter\@slowromancap\romannumeral #1@}
\makeatother
\begin{document}

\title{Non-linear charge and energy dynamics of an adiabatically driven interacting quantum dot} 

\author{Javier I. Romero}
\affiliation{International Center for Advanced Studies, ECyT-UNSAM,
  Campus Miguelete, 25 de Mayo y Francia, 1650 Buenos Aires, Argentina}

\author{Pablo Roura-Bas}
\affiliation{Dpto de F\'{\i}sica, Centro At\'{o}mico Constituyentes, Comisi\'{o}n
Nacional de Energ\'{\i}a At\'{o}mica, CONICET, Buenos Aires, Argentina}

\author{Armando A. Aligia}
\affiliation{Centro At\'{o}mico Bariloche and Instituto Balseiro, Comisi\'{o}n Nacional
de Energ\'{\i}a At\'{o}mica, CONICET, 8400 Bariloche, Argentina}

\author{Liliana Arrachea} 
\affiliation{International Center for Advanced Studies, ECyT-UNSAM, 
Campus Miguelete, 25 de Mayo y Francia, 1650 Buenos Aires, Argentina}


\begin{abstract}
We formulate a general theory to study the time-dependent charge and energy transport of 
an adiabatically driven interacting quantum dot in contact to a reservoir for arbitrary amplitudes of 
the driving potential. We study within this framework the Anderson impurity model with a local ac gate voltage. We show that the exact adiabatic quantum dynamics of this
system is fully determined by the behavior of the charge susceptibility of the frozen problem. 
At $T=0$, we evaluate the dynamic response functions with the numerical renormalization group (NRG). 
The time-resolved heat production exhibits a pronounced feature described by an instantaneous Joule law 
characterized by an universal resistance quantum $R_0=h/(2  e^2)$ for each spin channel. 
We show that this law holds  in non-interacting as well as in the 
interacting system and also when the system is spin-polarized. In addition, in the presence of a static magnetic field, 
the interplay between many-body interactions and
spin polarization leads to 
a non-trivial energy exchange between electrons with different spin components.
\end{abstract}

\date{\today}

\maketitle


\section{Introduction}

The generation of electron currents by locally applying time-dependent
voltages in coherent conductors is a topic of intensive research activity
for some years now. Any mechanism to be implemented with this goal is
accompanied by energy dissipation.

Quantum capacitors are prominent experimental realizations of these systems.
\cite{qcap1,qcap2,qcap3} They were introduced by B\"uttiker, Thomas and
Pr\^etre as quantum equivalents of the classical linear RC 
circuits, \cite{btp1,btp2,btp3} by assuming a small amplitude of the driving voltage. The
corresponding ac complex impedance depends on the driving frequency, the
capacitance of the quantum dot and the resistance of the circuit. In the
original theory, \cite{btp1,btp2,btp3} transport coherence is assumed along
the full setup, and the only resistive element is the contact, which results
in a quantized electron relaxation resistance $R_q=R_0/N_c$
where $N_c$ is the number of  transport channels and  $R_0=h/(2  e^2)$, is
the resistance quantum. The universality of this resistance remains robust
in the low frequency regime upon adding electron-electron interactions in
the quantum dot provided that the system behaves as a Fermi liquid (FL). \cite%
{rosa1,rosa2,mora,miche1,miche2}

While in some experiments the driving amplitudes were within the range of
linear response theory, \cite{qcap1} further experimental \cite{qcap2,qcap3}
and theoretical \cite{nonlin,misha1,misha2,misha3} contributions focused on
quantum capacitors as single-electron sources, implying large amplitudes. In
Ref. ~\onlinecite{nonlin} a theory for the regime of large amplitudes was
proposed for non-interacting systems. The effect of many-body interactions
was later considered within perturbation theory, \cite{janine} mean-field
approximations, \cite{david} and exact approaches valid in the large-transparency limit.\cite{lbf} 
One of the goals of the present contribution is
to study the low-frequency non-linear regime while fully taking into account
many-body interactions and spin-polarization effects caused by external magnetic fields.

\begin{figure}[tbp]
\includegraphics[width=.7\columnwidth]{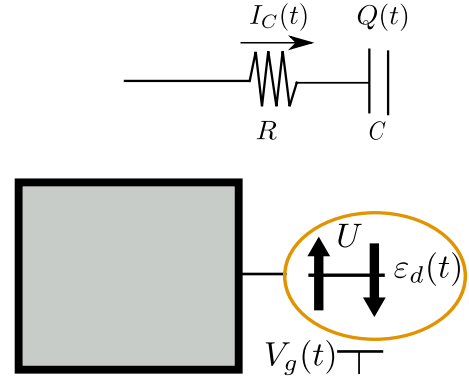}
\caption{(Color online) Sketch of the setup. A quantum dot described by a
single electron level with Coulomb interaction $U$ and   is driven by an ac gate
voltage $V_g(t)= V_0 \sin (\Omega t)$ and is connected to a normal lead. Top: representation of the setup 
in terms of a resistance connected in series with a capacitor.}
\label{fig1}
\end{figure}
The setup consists of a quantum dot driven by a gate voltage $V_g(t)$ and
connected to an electron reservoir, as sketched in Fig.1. 
We focus on the so called adiabatic regime where  the time scale associated to a 
variation of $V_g(t)$ is much larger than the characteristic time scale for the
dynamics of the electrons inside the quantum dot. 

As mentioned before, in linear response, it is usual to represent 
this setup in terms of a resistance in series with a capacitor, 
as sketched at the top of Fig.1. 
\cite{qcap1,qcap2,qcap3,btp1,btp2,btp3,rosa1,rosa2,mora,miche1,miche2} In this paper, we show that this representation  with $R=R_q$ is also sound to describe the adiabatic dynamics of the interacting system without magnetic field beyond linear response.
In the case of a magnetic field applied at the quantum dot, we analyze the setup in the context of the circuit sketched 
in Fig. \ref{circuit2}, where
each spin channel is regarded as a branch of
a circuit with a capacitance  in series with a resistance  accounting for  a
total voltage drop $V_g(t)$.  
For the quantum dot without many-body interactions, we show that the resistance per spin channel is 
$R_0$, while we argue that for the interacting quantum dot with magnetic field,
the charge dynamics cannot be properly represented by this circuit.

The quantum dot-reservoir system dissipates the energy received from the ac source in the form of heat 
that flows into the reservoir. For a
non-interacting quantum dot connected to a single-channel  reservoir at zero
temperature and for low frequency but arbitrary amplitude of the driving
potential, the time-dependent rate for the dissipation of energy was found
to obey an instantaneous Joule  law (IJL), with  the universal resistance 
$R_0$,  \cite{Ludovico2014,ludo16}
\begin{equation}  \label{ijl}
P_{\mathrm{Joule}}(t)= R_0 \sum_{\sigma} [I_{C,\sigma}(t)]^2,
\end{equation}
where $I_{C,\sigma}(t)$ is the instantaneous charge current of electrons with spin $\sigma$ flowing 
from the quantum dot to the reservoir.
Here we investigate to what an extent the Coulomb interaction at the quantum dot 
affects this picture. We analyze this ingredient in the framework of the Anderson impurity model. 
One of the scenarios in this context is the Kondo effect, which takes place
below the so called
Kondo temperature $T_K$ when the quantum dot is strongly connected to the
reservoirs and occupied by an odd number of electrons. \cite{kondo} The
electrons of the reservoir and the  effective spin 1/2 localized at the
quantum dot form a many-body singlet state.  Other scenario is  the Coulomb blockade, according to which 
 it is necessary to overcome the energy of the Coulomb interaction to introduce an additional electron in the quantum dot once it is already filled with one electron.
 In all the regimes, the single impurity Anderson model  behaves as a FL, even in the presence of a magnetic field. We show that, due to this fact, the dynamics for the 
 energy dissipation in the adiabatic regime is ruled by the  IJL of  Eq. (\ref{ijl}) even beyond linear response. 
 However, the mechanisms for the energy transport 
depend on the interactions and the spin polarization.
We show that in systems without spin polarization (interacting and non-interacting), 
as well as in non-interacting systems (with and without spin-polarization), 
electrons with each spin orientation separately dissipate energy at a rate described by a Joule law 
$P_{ \rm{Joule}, \sigma} (t)= R_0 [I_{C, \sigma}(t)]^2$. Instead, the interplay between many-body interactions 
and spin polarization leads to regimes where electrons with a given spin orientation 
exchange energy with electrons with the opposite spin orientation, although the total rate for the energy 
dissipation is described by Eq. (\ref{ijl}).

The work is organized as follows. We present the theoretical treatment in Section II. In Section III we 
discuss the case where the quantum dot is non-interacting. We show that the exact description of the adiabatic dynamics is fully determined by the
behavior of the charge susceptibility of the frozen system described by the equilibrium Hamiltonian frozen at a given time.
The effect of many-body interactions is discussed in Section IV.  In Section V we present numerical results 
obtained with Numerical Renormalization Group (NRG). 
For systems without spin polarization, we also use exact results of static properties obtained using the
Bethe Ansatz (BA). We present the summary and conclusions in Section VI. 

\section{Theoretical treatment}

\subsection{Model}

We consider the system of Fig. \ref{fig1}. A driven quantum dot is
connected to a normal lead of free electrons at zero temperature and
chemical potential $\mu$. The full setup is described by an Anderson
Hamiltonian, 
\begin{equation}  \label{Hamtot}
H(t)= H_{\text{dot}}(t)+H_{\text{res}} + H_{\text{T}}.
\end{equation}
The first term describes the dot 
\begin{equation}  \label{ham}
H_{\text{dot}} (t) = \sum_{\sigma} \varepsilon_{d, \sigma}(t)n_{d\sigma} + U
\left(n_{\uparrow}-\frac{1}{2}\right)\left(n_{\downarrow}-\frac{1}{2}\right),
\end{equation}
with $n_{d\sigma}$ denoting the number operator with spin $\sigma=\uparrow,\downarrow$, 
$U$ is the Coulomb repulsion, and $%
\varepsilon_{d,\sigma}(t)=\varepsilon_{0}+ s_{\sigma} \frac{\delta_Z}{2}+
{\cal V}_{g}(t)$ is the single-particle energy modulated by the applied
gate voltage $V_g(t)$, with ${\cal V}_{g}(t)= e V_g(t)= V_0 \sin (\Omega
t)$,  $\delta_Z$ is the Zeeman splitting due to the presence of an
external magnetic field, $s_{\sigma}= \pm 1$ for $\sigma= \uparrow, \downarrow$, 
and $-e$ is the charge of the electron. The reservoir is described by the Hamiltonian $H_{\text{res}}
=\sum_{\sigma,k}\epsilon_{k}c_{k\sigma}^{\dagger}c_{k\sigma}$, which is
assumed to have a constant density of states within a bandwidth $2D$. 
The  coupling between dot and reservoir is 
$H_{T}=V_{c}\sum_{k\sigma}\left[c_{k\sigma}^{\dagger}d_{\sigma} +h.c\right]$.

\subsection{Charge and energy adiabatic dynamics}
The conservation of the charge in the full system implies 
\begin{equation}
e\dot{n}_{d}(t)=e\sum_{\sigma }\dot{n}_{d\sigma }(t)=\sum_{\sigma
}I_{C,\sigma }(t),  \label{ic}
\end{equation}
where $n_{d\sigma }(t)\equiv \langle n_{d\sigma }(t)\rangle $ is the occupancy of
the dot  by electrons with spin $\sigma $ at time $t$,  
$I_{C,\sigma }(t)$ is the contribution of the electrons with spin $\sigma$
to the charge current flowing out of the dot at time $t$, and $e > 0$ the elementary charge.

The power developed by the external ac source on the electron system is defined as \cite{exch} $P_{\rm ac} (t) = - \langle  \partial H /\partial t \rangle = -e \sum_{\sigma }n_{d\sigma }(t)\dot{V}_{g}(t)$.
This  leads to a net heat production  in the electron system at a rate  $\dot{Q}(t)= -P_{\rm ac}(t)$. \cite{ludo16}  
We find it convenient to define the power 
  \begin{equation}
P(t)=e\sum_{\sigma }n_{d\sigma }(t)\dot{V}_{g}(t), \label{pac}
\end{equation}
such that  $P>0$ implies work delivered from  the electron system against the ac sources. With this definition, the rate for the heat production in the electron system reads $\dot{Q}(t)= P(t)=P_{\mathrm{cons}}(t)+P_{\mathrm{diss}}(t)$.  \cite{note}
This 
power contains a purely ac  component $P_{\mathrm{cons}}(t)$ associated to the reversible heat produced 
by the conservative (Born-Oppenheimer) forces,
and a dissipative component $P_{\mathrm{diss}}(t)$ with a non-zero time
average.

The dynamics of the heat production and the charge current is fully
determined by $n_{d\sigma }(t)$. For low frequencies, the latter can be
calculated within the adiabatic formalism of Ref. ~\onlinecite{adia}, which
corresponds to linear-response in $\dot{V}_{g}(t)$ (see Appendix A). The result is 
\begin{equation}
n_{d\sigma }(t)=n_{f\sigma }(t)+ e \Lambda _{\sigma }(t)\dot{V}_{g}(t),
\label{nd}
\end{equation}%
where $n_{f,\sigma }(t)\equiv \langle n_{d\sigma }\rangle _{t}$ is the
snapshot occupancy of the dot, evaluated with the exact \emph{equilibrium}
density matrix $\rho _{t}$ corresponding to the Hamiltonian $H(t)$ frozen at
the time $t$. The coefficient of the second term is
\begin{equation}
\Lambda _{\sigma }(t)=- \lim_{\omega \rightarrow 0}\frac{\mbox{Im}\lbrack 
\chi_{t}^{\sigma  \sigma }(\omega ) +  \chi_{t}^{\sigma  \overline{\sigma} }(\omega ) ]}{\hbar \omega },  
\label{lambda}
\end{equation}%
with $\overline{\uparrow}= \downarrow$ and $\overline{\downarrow}= \uparrow$.
$\chi _{t}^{\sigma \sigma^{\prime} }(\omega )$ is the Fourier transform of the charge
susceptibility $\chi _{t}^{\sigma \sigma^{\prime} }(t-t^{\prime })=-i\theta (t-t^{\prime
})\langle \lbrack n_{d\sigma }(t),n_{d\sigma^{\prime} }(t^{\prime })]\rangle _{t}$
evaluated with $\rho _{t}$.

In the case of the system with applied magnetic field, it is appropriate to analyze separately the current and the power developed by
electrons with the different spin components. 
The current per spin can be calculated by the derivative of Eq.~(\ref{nd}) 
\begin{equation}
I_{C,\sigma }(t)=e\frac{dn_{f,\sigma }}{dV_{g}}\dot{V}_{g}(t)+e^{2}\frac{d
\left[ \Lambda _{\sigma }(t)\dot{V}_{g}(t)\right] }{dt},  \label{cur}
\end{equation}
where the first term is related to the static charge susceptibility through
$dn_{f,\sigma }/dV_{g} = \chi_t^{\sigma \sigma}(0)$. 

The frozen component $n_{f,\sigma}(t)$ contributes to the conservative component of this power, 
while the last term of Eq.~ (\ref{nd}) contributes  to the
non-conservative one. They read, respectively
\begin{equation}
P_{\mathrm{cons}, \sigma}(t) = e  n_{f \sigma} (t) \dot{V}_g(t), 
\;\;\;\;\;\;P_{\sigma}(t) =  e^2 \Lambda_{\sigma}(t) [\dot{V}_g(t)]^2.
\label{psi}
\end{equation}
  It is important to notice that the non-conservative components $P_{\rm \sigma}(t)$ are not necessarily 
  fully dissipative. They certainly contribute to the total dissipation, but they may also
  contain a non-dissipative "exchange" part $P_{\rm ex}(t)$, such that 
  $P_{\uparrow (\downarrow)}(t)= \pm P_{\rm ex}(t) + P_{\rm diss, \uparrow (\downarrow)}(t)$. The exchange component is associated to time-dependent induced forces that
  are proportional to $\dot{V}_g(t)$. In this sense, these forces are akin to 
  the "Lorentz" forces discussed in Ref. \onlinecite{lu}. However, in the present case they may develop work only instantaneously while the average over one period is zero.

The total power has conservative $P_{\rm cons}(t)= \sum_{\sigma} P_{\mathrm{cons}, \sigma}(t)$, 
and dissipative components $P_{\rm diss}(t)= \sum_{\sigma} P_{\mathrm{diss}, \sigma}(t)$, which read
\begin{eqnarray}  \label{pacad}
P_{\mathrm{cons}}(t) &=& e \sum_{\sigma} n_{f \sigma} (t) \dot{V}_g(t),  \notag \\
P_{\mathrm{diss}}(t) &= & e^2 \sum_{\sigma} \Lambda_{\sigma}(t) [\dot{V}_g(t)]^2.
\end{eqnarray}

For later use we also define
\begin{eqnarray}
\Lambda _{\sigma \sigma^{\prime}}(t)=- \lim_{\omega \rightarrow 0}\frac{\mbox{Im}\lbrack 
\chi_{t}^{\sigma  \sigma^{\prime} }(\omega ) ]}{\hbar \omega }, \notag \\
P_{\sigma \sigma^{\prime}}(t) = e^2 \Lambda_{\sigma \sigma^{\prime}}(t) [\dot{V}_g(t)]^2.
\label{lambdass}
\end{eqnarray}

When performing the averages over one period $\tau= 2 \pi/\Omega$ for these
two contributions to the power,  $\overline{P}_{\mathrm{cons, diss}}=
(1/\tau) \int_0^{\tau} dt P_{\mathrm{cons, diss}}(t)$, we find
$\overline{P}_{\mathrm{cons}}=0$ and $\overline{P}_{\mathrm{diss}} \geq 0$ in accordance to the second law of thermodynamics. 

We see that the full charge and energy dynamics in the adiabatic regime is completely determined by the behavior of the frozen charge susceptibility $\chi_t^{\sigma \sigma^{\prime}}(\omega)$, 
irrespective  of the strength
of the interactions and the amplitude of the driving potential.

\begin{figure}[tbp]
\includegraphics[width=.7\columnwidth]{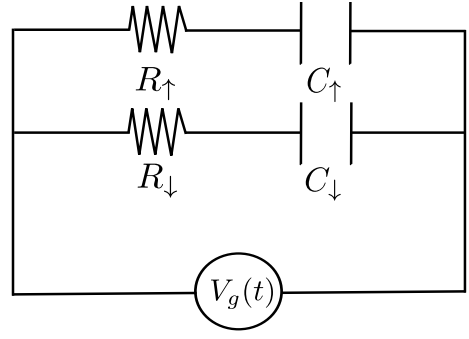}
\caption{(Color online) Sketch of the circuit. Upper and lower branch
corresponds to $\uparrow$ and $\downarrow$ spin channels.}
\label{circuit2}
\end{figure}

\subsection{Analogy to the non linear classical circuit}
We now discuss the representation of the equations for the dynamics of the
charge and energy introduced in the previous section, in terms of a
classical non-linear circuit. We find it convenient to treat the two
spin channels separately as a circuit with two branches (one for each spin
species) connected in parallel to the ac source, as sketched in Fig. \ref{circuit2}. 
Each branch contains a  capacitance $C_{\sigma }(t)$ in series with a resistance 
$R_{\sigma }(t)$. 

We assume that the equation relating the current through each branch of the circuit with the potential $V_g(t)$ is
\begin{equation}
I_{C,\sigma }(t)= -C_{\sigma }(t)\dot{V}_g(t)+ 
e^2 \frac{d\left[R_{\sigma }(t) C_{\sigma }(t)^{2}\dot{V}_g(t) \right]}{dt}.  \label{circuit}
\end{equation}
As discussed in Appendix \ref{eqcirc}, this equation corresponds to a true macroscopic classical RC circuit 
in the non-linear low-frequency regime satisfying 
$\Omega R_{\sigma} I_{\sigma} \ll 1$. 
Identifying linear and quadratic terms in $\dot{V_{g}}$ in the above equation with those of the quantum current Eq. (\ref{cur}) 
one obtains 
\begin{equation} 
C_{\sigma }(t)=-e\frac{dn_{f,\sigma }(t)}{dV_{g}}= -e  \chi_t^{\sigma \sigma}(0),\;\;\;R_{\sigma }(t)C_{\sigma }^{2}(t)=e^{2}\Lambda _{\sigma }(t).  \label{cr}
\end{equation}
Here, unlike the linear case, the non-linear capacitance $C_{\sigma }(t)$
and resistance $R_{\sigma}(t)$ are, in general, functions of $t$. In terms of these
coefficients, the dissipated power (\ref{pacad}) reads 
\begin{equation}
P_{\mathrm{diss }}(t)= \sum_{\sigma} R_{\sigma }(t)C_{\sigma }^{2}(t)[\dot{V}%
_{g}(t)]^{2}.  \label{disipa}
\end{equation}%
We see that for the case where $R_{\sigma }(t)=R_{0}$, Eq. (\ref{disipa})  reduces to the IJL
described by Eq. (\ref{ijl}), which is immediately derived by retaining only
the first term of Eq. (\ref{cur}). In fact, the latter is the only term of
Eq. (\ref{cur}) which has a contribution $\propto \lbrack \dot{V}_{g}(t)]^{2}
$ to $P_{\mathrm{diss }}(t)$, since $d\Lambda _{\sigma }(t)/dt=
\left(d\Lambda _{\sigma }(t)/dV_{g}(t)\right) \dot{V_{g}}(t)$.

The above equations are exact and valid in general within the adiabatic regime. However, 
in order to establish a meaningful correspondence between  
the charge and energy dynamics of the quantum system and the classical circuit of Fig. \ref{circuit2}, 
the coefficients defined in Eq. (\ref{cr}) should
 also verify  $R_{\sigma}(t) >0$ and  $C_{\sigma}(t) >0$.  
As we will discuss in detail in the next sections, such a correspondence  is valid 
in the system  without spin polarization ($\delta_Z=0$) for arbitrary  Coulomb interaction $U$, 
as well as in
the non-interacting case ($U=0$) with arbitrary $\delta_Z$. 
We also argue in Sections IV and V that the dynamics of the
driven interacting quantum dot in the presence of a magnetic field  
cannot be interpreted in terms of the circuit of Fig. \ref{circuit2}.

\section{ Non-interacting electrons}
\subsection{Review of the spinless case}
The expressions of the previous sections are completely general and valid for arbitrary
temperatures, for non-interacting as well as interacting systems. We now
relate them to the non-interacting results for spinless electrons of Refs. ~
\onlinecite{nonlin,Ludovico2014,ludo16,adia}. This corresponds to the
Hamiltonian (\ref{ham}) with $U=\delta _{Z}=0$ and with only one spin
species. Following Refs. ~\onlinecite{lili,Ludovico2014,ludo16}, we get 
\begin{eqnarray}
C(t) &=&-e\int d\epsilon \frac{\partial f}{\partial \epsilon }\rho
_{f}(t,\epsilon ),  \notag \\
\Lambda (t) &=&- \frac{h}{2}\int d\epsilon \frac{\partial f}{\partial \epsilon 
}\left[ \rho _{f}(t,\epsilon )\right] ^{2},  \label{noni}
\end{eqnarray}%
where  $\rho_{f}(t,\epsilon )=(\Delta/\pi) /\left[ \left( \epsilon -\varepsilon
_{d}(t)\right) ^{2}+\Delta ^{2}\right] $ is the non-interacting frozen
density of states of the quantum dot connected to a reservoir with constant
density of states $\nu $, $\Delta =\pi \nu V_{c}^{2}$, and $f(\epsilon )$
is the Fermi distribution function.

The resistance can be directly calculated from Eq. (\ref{cr}). At $T=0$, we
have $-\partial f/\partial \epsilon =\delta (\epsilon -\mu )$. Hence, the
coefficients simplify to $C(t)=e^{2}\rho _{f}(t,\mu )$ and $%
R(t)=R_{0}=h/(2e^{2})$. The latter corresponds to the universal resistance
quantum for a single channel. By substituting these expressions in 
$P_{\mathrm{diss}}(t)$ and $I_{C}(t)$, and keeping terms up to 
$\mathcal{O}(\dot{V}_{g}^{2})$, we recover the IJL of Eq. (\ref{ijl}) as in Ref.~%
\onlinecite{Ludovico2014}. Interestingly, we get the same expression of the
current $I_{C}(t)$ in the non-interacting limit as the one of Ref. ~%
\onlinecite{nonlin}. However the definition of $R(t)$ presented there
differs from the definition of Eq. (\ref{cr}) with $\Lambda (t)$ given by 
Eq. (\ref{noni}). Such difference should be traced back to the equation for the
non-linear circuit (\ref{circuit}). Unlike the one considered in Ref. ~%
\onlinecite{nonlin}, Eq. (\ref{circuit}) includes the factor $RC$ inside the
time-derivative of the second term. The structure of the latter Eq. is
motivated by the adiabatic expansion of the occupancy Eq. (\ref{nd}), by
identifying the coefficient $\Lambda (t)$ as the dissipative contribution.
Remarkably, our definition of $R(t)$ can be easily related to $R_{0}$ in the
limit of $T=0$ and it consistently leads to the Joule law of Eq. ~(\ref{ijl}%
), while it is also in agreement with the effective resistance defining the
noise. \cite{nonlin}

\subsection{Spinfull electrons}
We now consider the case with $U=0$ and arbitrary $\delta_Z$. 
Notice that for non-interacting electrons the "crossed-susceptibility" 
$\chi_t^{\sigma,\overline{\sigma}}(\omega)=0$. Hence the  coefficient $\Lambda_{\sigma}(t)$  
is fully determined 
by the susceptibilities $ \chi_t^{\sigma,\sigma}(\omega)$. The calculations of 
Refs. \onlinecite{nonlin,Ludovico2014,ludo16,adia} can be easily extended to non-interacting 
electrons with spin.

The frozen occupancy of the quantum dot with spin $\sigma$ is
\begin{equation}
n_{f,\sigma}(t)= \int d \epsilon \rho_{f,\sigma}(t,\epsilon) f({\epsilon}),
\end{equation}
where $\rho_{f,\sigma}(t,\epsilon)= (\Delta_{\sigma}/\pi) /\left[ \left( \epsilon -\varepsilon
_{d,\sigma}(t)\right) ^{2}+\Delta_{\sigma}^{2}\right] $.
For this model $\partial  \rho_{f,\sigma}(t,\epsilon) /\partial t =
 e \left( \partial \rho_{f,\sigma}(t,\epsilon) /\partial \epsilon \right) \dot{V}_g(t) $. 
Hence, after integrating by parts the above equation, we get
for $T=0$
\begin{equation} \label{capnonint}
C_{\sigma}(t)= e \rho_{f, \sigma}(t, \mu).
\end{equation}
In addition, we get an expression like (\ref{noni}) for each spin orientation $\sigma$. 
For $T=0$, it reads
\begin{equation}\label{lamnons}
\Lambda _{\sigma }(t)=  \frac{h}{2} [ \rho_{f,\sigma}(t,\mu) ]^2 = 
  \frac{h}{2} [\chi^{\sigma \sigma}_t(0) ]^2,
\end{equation}
which is a special case of the Korringa-Shiba (KS) law discussed in the next section. 
Inserting these expressions in Eqs. (\ref{cr}) we obtain
$R_{\sigma}(t)=R_0$. Substituting  in (\ref{pacad}), we see that the dissipated power is ruled by the IJL of 
Eq. (\ref{ijl}). 

Therefore, for non-interacting electrons, Korringa-Shiba law of 
Eq. (\ref{lamnons}) implies that there is a full one-to one correspondence between the charge and energy dynamics 
of the driven electron system and the two-branch 
circuit sketched in  Fig. \ref{circuit2}, with 
resistances $R_{\sigma}(t)=R_0$, 
even when the electrons are spin-polarized. This also means that 
the ac forces associated to the induced charge for each spin orientation dissipate 
heat in the form of a Joule law, $P_{ \sigma}(t)= R_0 [I_{C, \sigma}(t)]^2=P_{\rm Joule, \sigma}(t)$. Hence,
 $P_{\rm diss}(t)= \sum_{\sigma} P_{\rm Joule, \sigma }(t)= R_q [I_C(t)]^2$, with $I_C(t)= \sum_{\sigma} I_{C,\sigma}(t)$ and $R_q= R_0/2$.

\section{Interacting electrons}
\subsection{Exact results}
For interacting electrons, the crossed susceptibility $\chi_t^{\sigma,\overline{\sigma}}(\omega)$ 
contributes to the coefficient $\Lambda_{\sigma}(t)$, in addition to
$\chi_t^{\sigma,\sigma}(\omega)$. For Fermi liquids, an important relation  exists for the total 
charge susceptibility $\chi_t^c(\omega)= \sum_{\sigma, \sigma^{\prime}}
 \chi_t^{\sigma,\sigma^{\prime}}(\omega)$, which receives the name of Korringa-Shiba law. \cite{korr-shiba} 
 In the non-interacting case, it is expressed in Eq. (\ref{lamnons}). 
In the interacting case, it is a non-trivial result, which  was originally proved by Shiba  
in the Anderson model \cite{korr-shiba}  and later generalized by Fillipone {\it et al.} when a
magnetic field is also considered. \cite{miche1,miche2}
 It reads
 \begin{equation} \label{ks}
 \lim_{\omega \rightarrow 0} \frac{ \mbox{Im}\left[ \chi^c_t(\omega) \right]}{\hbar \omega} 
 = - \frac{h}{2} \sum_{\sigma}  [\chi^{\sigma \sigma}_t(0) ]^2.
 \end{equation}
  This relation has been used to study the present problem within the linear response 
  regime. \cite{rosa1,rosa2,mora,miche1,miche2}
Here, we show that the Korringa-Shiba law Eq. (\ref{ks}) is equivalent to the instantaneous Joule 
law Eq. (\ref{ijl}), even in the presence of a magnetic field and also in the non-linear response regime. 

In fact, from Eqs. (\ref{lambda}) and (\ref{pacad}) and taking into account that Eq. (\ref{ks}) is satisfied, we have 
\begin{equation}\label{pdis}
P_{\rm diss}(t)= \frac{ e^2  h}{2}  \sum_{\sigma}  [\chi^{\sigma \sigma}_t(0) ]^2 \dot{V}_g(t)^2
\end{equation}
On the other hand, the -up to ${\cal O}(\dot{V}_g(t))$- charge current with spin $\sigma$ 
is given by the first term of Eq. (\ref{cur}) and reads
\begin{equation}
I_{C, \sigma}(t)\simeq e \frac{\partial n_{f \sigma}(t)}{\partial V_g(t)} \dot{V}_g(t) =  
e \chi^{\sigma \sigma}_t(0)\dot{V}_g(t).
\end{equation} 
Then, substituting in Eq. (\ref{pdis}), we get
\begin{equation}
P_{\rm diss}(t)= \frac{h}{2 e^2} \sum_{\sigma} \left[ I_{C,\sigma}(t) \right]^2, \label{ijl1}
\end{equation}
which is, precisely, the IJL. 
This result holds for electrons with and without spin polarization, in the non-linear as well as in 
the linear regimes.

\subsection{Non-polarized electrons and the representation by the classical circuit}
In the case of non-polarized electrons, where the two spin orientations are equivalent.
The total charge current $I_C(t)=\sum_{\sigma}I_{C,\sigma }(t)$ associated to the change in the dot 
occupancy by up and down spins
is given by [see Eqs. (\ref{cur}) and (\ref{cr})]
\begin{equation}
I_C(t)= -C(t)\dot{V}_g(t)+ e^2 \frac{d\left[ \Lambda^c(t) \dot{V}_g(t) \right]}{dt},  \label{circuitt}
\end{equation}
with $C(t)= \sum_{\sigma} C_{\sigma}(t)$, and  $\Lambda^c(t)= \sum_{\sigma} \Lambda_{\sigma }(t)$.

The definition of the resistance  given in Eq. (\ref{cr}), along with 
$C(t)= -e^2 \chi_t^{c} (0)$
and the KS relation Eq. (\ref{ks}), lead to 
the resistance $R_0$ for each spin channel or an equivalent resistance $R_q=R_0/2$ for the 
equivalent circuit of Fig.1.
In fact,
because of the equivalence of both branches of the  circuit of Fig. \ref{circuit2}, 
the voltage drop at the middle point between the resistance and capacitance for each
branch are the same. Therefore,  one can connect this two points with a cable carrying no current 
and the circuit becomes equivalent 
to that at the top of Fig. 1, with an effective resistance  $R_q^{-1}= 2 R_0^{-1}$ in 
series with the effective capacitance $C(t)=2 C_{\sigma}(t)$. In addition, the arguments of the next section as well as the numerical results will show that $C(t) \geq 0$.
Hence, for non-polarized electrons, the behavior of the charge and energy dynamics is consistent with 
the representation of the 
setup in terms of the parallel circuit of Fig. \ref{circuit2} or the equivalent one of Fig. 1.
The charge dynamics is described by Eq. (\ref{circuitt}) with $e^2 \Lambda^c(t)=R_q [C(t)]^2$.
while the dissipated power obeys the IJL Eq. (\ref{ijl1}).

\subsection{Polarized electrons in the random-phase approximation}
In the case of polarized electrons, the two spin orientations are not equivalent and it is not easy to analyze the dynamics by simple analogy to the classical circuit. 
 It is important to notice that the Coulomb interaction effectively renormalizes the gate voltage at a given time. At the mean field level, this can be accounted by an occupancy-dependent 
 term $U \langle n_{f, \overline{\sigma}} \rangle $
 which adds to ${\cal V}_g(t)$ in the effective local energy experienced by an electron with spin $\sigma$ at the quantum dot. 
Here, we will analyze the consequence of this effect on the basis  of the behavior of 
 the charge susceptibility in the "random phase approximation" (RPA). In the next section we will present a more accurate  analysis based on NRG results.

RPA corresponds to calculating the dynamic susceptibility from the summation of an infinite perturbative series of "bubble" diagrams. The result in the present case is 
\begin{equation}\label{rpa}
\chi_t^{\sigma \sigma}(\omega) = \frac{ \chi_t^{0 \sigma}(\omega) \left[1 + U  \chi_t^{ 0 \overline{\sigma}}(\omega) \right]}{1- U^2 \chi_t^{0 \sigma}(\omega)  \chi_t^{ 0 \overline{\sigma}}(\omega)}, 
\end{equation}
where $\chi_t^{0 \sigma}(\omega)$ is the susceptibility for $U=0$. The latter is a function of the gate 
voltage $V_g(t)$ and satisfies the KS relation (\ref{lamnons}). The static limit is given by 
Eq. (\ref{capnonint}), $\chi_t^{0 \sigma}(0)=-\rho_{f,\sigma}(\mu)$. 
In the case of non-polarized electrons, where the two spin orientations are equivalent, 
we have $\chi_t^{\sigma \sigma}(\omega) = \chi_t^{0 \sigma}(\omega)/\left[1+ U \chi_t^{0 \sigma}(\omega) \right]$. 
Hence,  the calculation of the capacitance gives 
$C_{\sigma}(t)= C_{\sigma, 0}(t) /\left[1+ U C_{\sigma, 0}(t)/e \right] $, where $C_{\sigma, 0}(t) $ 
is the capacitance of the non-interacting system (\ref{capnonint}), which indicates that $C_{\sigma}(t) \geq0$. 

In the case of polarized electrons
we have situations where $|\chi_t^{0 \sigma}(0)| \ll |\chi_t^{0 \overline{\sigma}}(0)|$ or vice versa, 
in which case Eq. (\ref{rpa}) leads to  negative values of the coefficient $C_{\sigma}(t)$ 
for large enough $U$. This corresponds to a current 
between the reservoir and the dot which opposes to the sense of circulation imposed by the voltage drop. Furthermore, after some algebra from (\ref{rpa}) and the KS relation for the non-interacting susceptibilities
(\ref{lamnons}), we can see that in such situations 
the signs of $\Lambda_{\sigma}(t)$ and $\Lambda_{\overline{\sigma}}(t)$ are opposite.
This would correspond to instantaneous exchange of power between the two spin species, 
$P_{\sigma}(t) \sim P_{\rm ex}(t) \sim -P_{\overline{\sigma}}(t)$. 
We can interpret this behavior as  electrons with spin $\sigma$  receiving energy from the electrons 
with spin $\overline{\sigma}$ to move against the
external voltage drop. In addition to this component, there is a dissipative component of the power 
satisfying the IJL [see Eq. (\ref{ijl1})]. 
Since the behavior explained above
is not expected in a capacitive circuit element, we conclude that the representation of the dynamics of the driven interacting quantum dot in the presence of a magnetic field is not properly represented by a 
circuit  like that of Fig. \ref{circuit2}. In the next section, we will verify that such a behavior indeed takes place when the susceptibilities are exactly calculated with NRG.

\section{Numerical results for the non-linear interacting regime}
We now turn to further analyze the adiabatic fully interacting case for arbitrary
amplitudes of the driving on the basis of numerical results. We use the numerical renormalization group (NRG)
algorithm of Ref. ~\onlinecite{Zitko2009} to compute the frozen occupancy
of the dot $n_{f\sigma }(t)$ and the charge susceptibility $\chi_{t}^{\sigma }(\omega )$. We stress 
that the evaluation of these two
quantities corresponds to an equilibrium calculation with the Hamiltonian 
$H(t)$ frozen at the time $t$ (for details see Appendix C). We also use analytical expressions 
of the impurity occupancy obtained from Bethe ansatz (BA) \cite{wt}
using the procedure outlined in the appendix of
Ref. \onlinecite{anda}.

\begin{figure}[tbp]
\centerline{
\includegraphics[width=0.9\columnwidth]{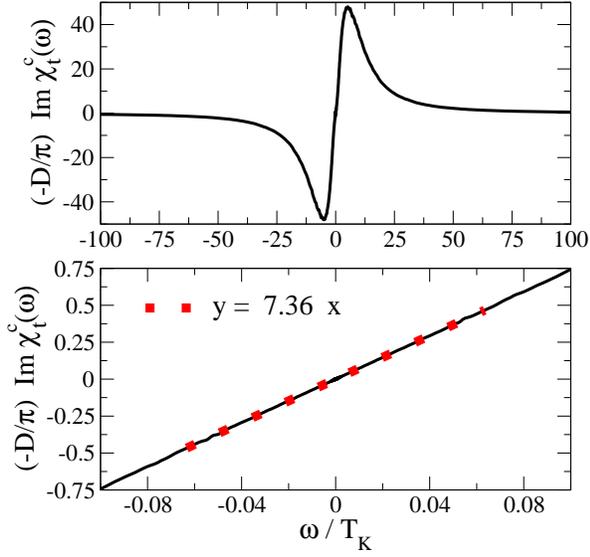}}
\caption{(Color online) Imaginary part of the dynamic susceptibility as a function of frequency
for $\Omega t= \pi/2$,  $\Delta=8x10^{-4}D$, $\protect\varepsilon_{0}=\protect\mu=0$, 
$U=0.05 D$, $V_0=0.024D$ and $T=0$.} 
\label{xi}
\end{figure}

\subsection{Results for zero magnetic field}
In a system without spin polarization, the two spin orientations are equivalent and the dynamics of the 
quantum system is fully equivalent to that of the 
circuit of Fig. \ref{circuit2} or that with the equivalent components sketched in Fig.1. As argued in the previous section, the two resistances 
can be simply substituted by the resistance $R_q=R_0/2$ in series with the capacitance 
$C(t)=2 C_{\uparrow}(t)=2 C_{\downarrow}(t)$. The total current circulating through the single branch of 
the equivalent circuit is given by Eq. (\ref{circuitt}).
The total dissipated power is given by Eq. (\ref{pacad}) with $\Lambda^c(t) = \sum_{\sigma} \Lambda_{\sigma}(t)$.

\subsubsection{Benchmark}
We start with a benchmark of the numerical results calculated with NRG against exact analytical results 
calculated with BA.
In particular, we verify that the Korringa-Shiba law Eq. (\ref{ks}) is satisfied. 
To this end, we analyze the Fourier transform of the total charge 
susceptibility 
$\chi_t^c (t-t^{\prime})= -i \theta(t-t^{\prime}) \sum_{\sigma \sigma^{\prime}} \langle [n_{d \sigma}(t), 
n_{d \sigma^{\prime}} (t^{\prime}) ]\rangle_t$  
evaluated with the exact equilibrium density matrix $\rho_t$. 
An example is shown
in Fig. \ref{xi} for a fixed time. For the parameters of the figure we estimate a Kondo temperature 
$T_K \approx 2.94 \times 10^{-4} D$,
where $D=1$ is half the band width used in the NRG calculations.
The general aspect of the curve is similar to that reported previously. \cite{rosa2}
According to the Korringa-Shiba law Eq. (\ref{ks})  one has for $\omega \rightarrow 0$ 
\begin{equation}
{\rm Im}\chi_t^c(\omega) =     - \omega\ \frac{h}{4}[\chi^c_{t}(0)]^{2} =  - \omega\ \frac{h}{2} \sum_\sigma[\chi _{t}^{\sigma }(0)]^{2}.
\label{fl2}  
\end{equation}

Fitting the results for $\omega \ll T_K$ as shown in the bottom panel of Fig. \ref{xi}, we obtain 
$-{\rm Im}\chi_t^c(\omega)/\pi = 7.36 \hbar^2  \omega\ /(D T_K)$,
or $-{\rm Im}\chi_t^c(\omega)= \Lambda^c \hbar \omega$, with
$\Lambda^c= 7.36 h /(2 T_K D)= 12520 h/D^2$. Calculating the charge susceptibility by numerical
differentiation of the total occupancy  $n_f(t)=\sum_{\sigma} n_{f \sigma}$, obtained with NRG from
\begin{equation} \label{chi0}
\chi^c_{t}(0)=   \sum_{\sigma} \frac{d n_{f \sigma}(t)}{d \varepsilon_d} ,
\end{equation}
with $\varepsilon_d\equiv \varepsilon_{d, \uparrow}(t)=  \varepsilon_{d, \downarrow}(t)$
 and 
using Eq. (\ref{fl2}), we obtain $\Lambda^c= 12732 h/D^2$, a value 1.7 \% larger.
By numerical differentiation of the BA occupancy we obtain $\Lambda^c= 12690 h/D^2$, which
differs from the previous result by 0.3 \%. This deviation might be due to the fact
that in the BA procedure we take $D \rightarrow \infty$. 

We have also checked the Fermi liquid relation Eq. (\ref{fl2}) for other values of the parameters
obtaining agreement with the static results within about 2 \%. This confirms the validity of these relations.
The slight discrepancy between both NRG results is likely to be due larger numerical errors 
in the dynamic calculation.

\begin{figure}[tbp]
\centerline{
\includegraphics[width=0.9\columnwidth]{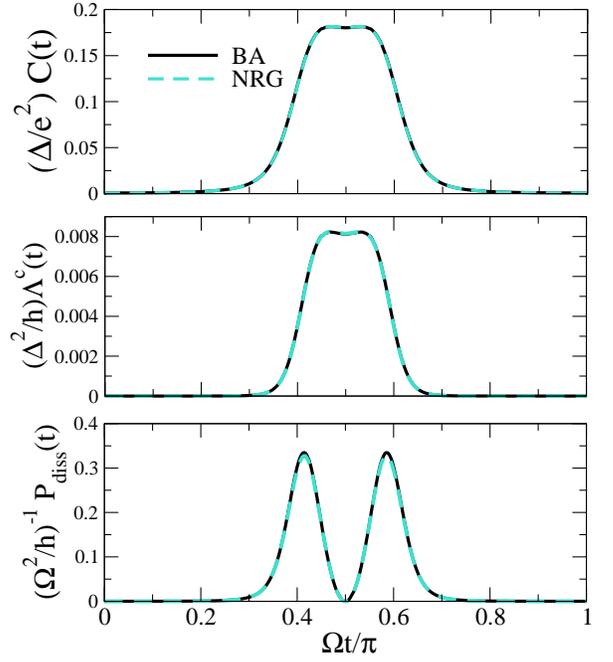}}
\caption{(Color online) (a) capacitance $C(t)$  (b) Dissipation coefficient $\Lambda^c(t)$,
and (c) dissipated power  $P_{\rm diss}(t)$ in the interacting non-linear
regime, as a function of time, calculated with two techniques.
Other parameters as in Fig. \ref{xi}.}
\label{clp}
\end{figure}
In Fig. \ref{clp} we represent the static susceptibility, proportional to the total 
capacitance $C(t)=- e^2 \chi_t^c(0)$ in the non-linear circuit analog, $\Lambda^c(t)$ and the total 
dissipated power $P_{\rm diss}(t)$
as a function of time. The total resistance is for all times $R_q=R_0/2$. 
The other parameters are the same as in the previous case, which was 
limited to $t= \pi/(2\Omega)$. One sees that the NRG and BA results agree 
very well for all times.

\subsubsection{Charge and energy dynamics}
From the practical point of view, it is easier to calculate static properties than dynamic ones. In addition, static quantities
can be calculated exactly with BA.  Hence, in what follows we calculate
$n_{f, \sigma}(t)$. Then, we calculate the static susceptibility from Eq. (\ref{chi0}). Finally, we use 
 the Korringa-Shiba relation Eq. (\ref{fl2})  to derive the dynamic response function $\Lambda^c(t)$. 

\begin{figure}[tbp]
\centerline{
\includegraphics[width=0.9\columnwidth]{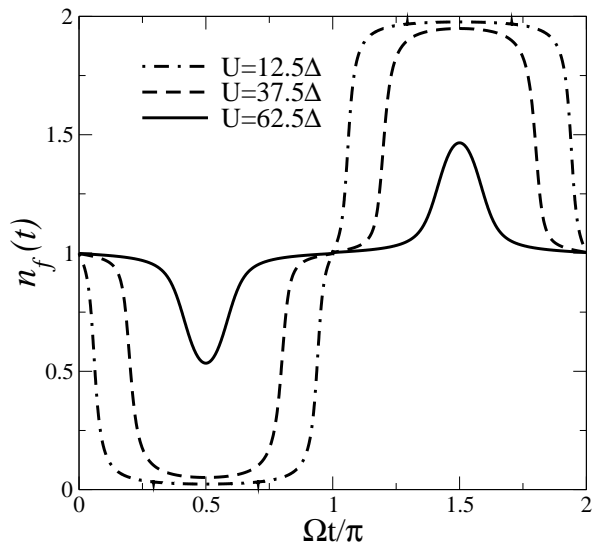}}
\caption{(Color online) occupancy of the quantum dot as a function of time for different values of 
the Coulomb interaction (indicated in the Fig). 
Other parameters as in Fig. \ref{xi}.}
\label{ocu}
\end{figure}

Results for the total frozen occupancy of the dot as a function of time, $n_f(t)$, are shown in Fig. \ref{ocu}.  
These correspond to a given driving amplitude and frequency, 
and different values of the Coulomb repulsion. To analyze these results, let us start by focusing on the plot with dashed-dot lines, corresponding to 
the smallest $U$.
At $t=0$ the dot is at the half-filling configuration, corresponding to a mean charge $n_f(0)=1$. 
As a function of $t$, 
$V_g$ 
increases 
and the occupancy of the dot decreases. In particular, when $\varepsilon_{d}(t) \sim U/2$, the quantum dot 
becomes empty and remains in that configuration while 
$V_g$
passes through its maximum at $\Omega t= \pi/2$. 
As the time continues to increase, the dot begins to get filled again with one electron. 
For larger time, $V_g$ continues to decrease and the single-particle energy of the dot 
$\varepsilon_{d}(t)$
reaches the value
$\sim -U/2$.  The dot gets filled with two electrons and remains at this occupancy as 
$V_g$ passes through its minimum at $\Omega t= 3 \pi/2$. Finally, the occupancy decreases
to reach the half-filled configuration, as the gate voltage completes the period. 
For the larger values of $U$ shown in the figure, we can observe similar features.  
For the largest one, corresponding to the
solid lines, the dot does not reach the occupancy with zero and two electrons, 
since for all times $|\varepsilon_{d}(t)| < U/2$.

\begin{figure}[tbp]
\centerline{
\includegraphics[width=0.9\columnwidth]{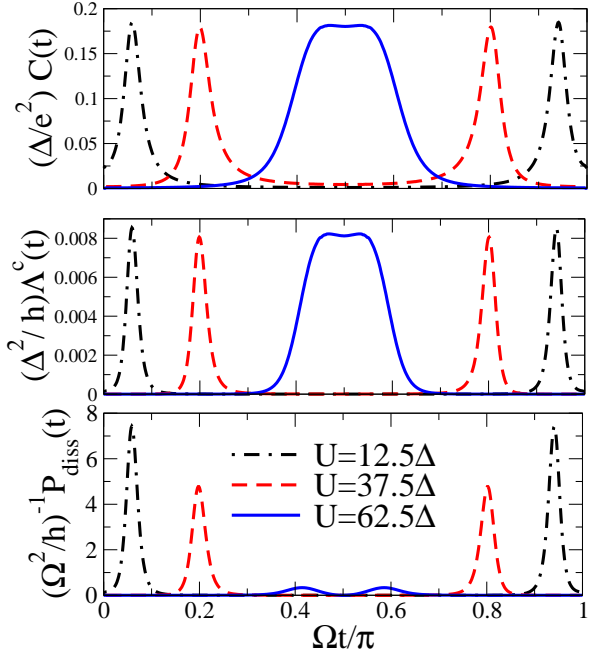}}
\caption{(Color online) Same as Fig. \ref{clp} calculated with BA for several values of $U$.} 
\label{clpu}
\end{figure}

The changes in the occupancy of the dot as a function of time, generate a charge current  
between the dot and the reservoir, according to Eq. (\ref{circuitt}). The corresponding capacitance, 
$C(t)$, the dissipation coefficient 
$\Lambda^c(t)$ and the
dissipated power $P_{\rm diss}(t)$ are shown in Fig. \ref{clpu}, for the same values of $U$ shown in Fig. \ref{ocu}.  We can identify features in the capacitance associated to the static charge susceptibility (\ref{chi0}).
Due to the Korringa-Shiba law, similar features are also found in the dynamic coefficient $\Lambda^c(t)$. This determines also the behavior of the dissipated power shown in the lowest panel of the figure, where we
can distinguish the peaks corresponding to the IJL described by Eq. (\ref{ijl}).

To close this section, we comment on the relation between the features characterizing the charge and energy dynamics and the behavior of the local frozen density of states at the quantum dot as a function of time.
 For gate voltages close to the symmetric configuration satisfying $|\varepsilon_{d}(t)| \sim 0$  there is one electron in the quantum dot and
the density of states has typically  a
resonant peak at the Fermi energy (the Kondo resonance) and charge-transfer (or Coulomb-blockade) peaks at high energies $\sim \pm U/2$. \cite{kondo}
As the gate voltage moves away from the symmetric configuration within the range $|\varepsilon_{d}(t) | \leq U/2$ the dot remains filled with a single electron, the Kondo resonance persists at the Fermi energy while
the  high-energy peaks move rigidly following $\varepsilon_{d}(t)$ (details of the evolution of the spectral weight can be found 
in e.g. Ref. \onlinecite{arf}). When $\varepsilon_{d}(t)\sim \pm U/2$, one of these peaks becomes 
aligned with the Fermi energy of the reservoir and the dot changes its occupancy to
 $0$ or $2$ electrons for $\varepsilon_{d}(t)= \pm U/2$, respectively. At the time this happens, a current flows towards or from the reservoir, respectively. This exchange of charge between the dot and the reservoir is accompanied 
 by an instantaneous dissipation of energy in the form of a Joule law, as described by Eq. (\ref{ijl}). This is reflected in the peaks of $C(t)$, $\Lambda^c(t)$ and $P_{\rm diss}(t)$ shown in  Fig. \ref{clpu}.
 
 In all the processes  discussed, the Kondo resonance does not play any significant role. Therefore, the behavior of Figs. (\ref{ocu}) and (\ref{clpu}) is also representative of the Coulomb blockade regime taking place at finite
 temperatures when the coupling to the reservoir is very weak. Interestingly, this behavior is also similar to what is observed in experiments 
of the compressibility of strongly correlated quantum dots, which are also related to the behavior of the charge 
susceptibility at the Fermi energy. \cite{compre}

\subsection{Results for finite magnetic field}
\begin{figure}[tbp]
\centerline{
\includegraphics[width=0.9\columnwidth]{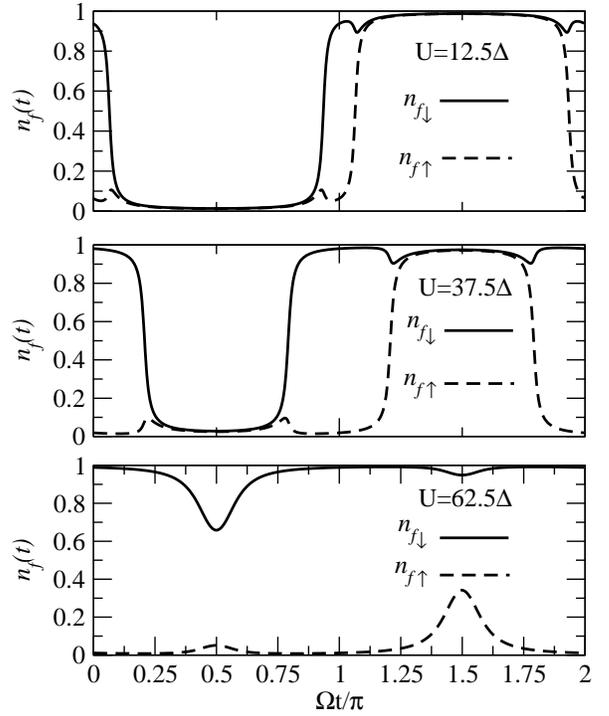}}
\caption{(Color online) Frozen occupancies $n_{f,\uparrow}(t)$ and $n_{f,\downarrow}(t)$ for a Zeeman splitting 
$\delta_Z=10^{-3}D$ and different values of  $U$. Other parameters are the same as in
the previous Figs.}
\label{nsig}
\end{figure}

The behavior of the time-dependent occupancies for $\uparrow$ and $\downarrow$ spins 
in the presence of a Zeeman splitting $\delta_Z$ at the quantum dot are shown in
Fig. \ref{nsig} for different values of $U$. The upper, middle and bottom panels correspond, 
respectively to the same parameters as the plot in dot-dashed, dashed and solid lines 
of Fig. 5, 
with the additional ingredient of a magnetic field. In all the cases, the dot is in
the half-filled configuration at $t=0$ ($n_f(0)=1$) and is predominantly occupied by electrons with spins 
parallel to the direction of the magnetic field (in this case $\downarrow$). 
The magnitude of the Zeeman splitting is chosen larger than the Kondo temperature $T_K$ for all times. 
Hence, for one electron in the dot, the system is in the local moment regime of the 
Anderson model with a significant spin polarization.
 
Starting from the symmetric configuration at $t=0$,  the energy of the localized electrons 
for both spin orientations $\varepsilon_{d,\sigma}(t)$
increases in time and the occupancy 
of the quantum dot evolves to the empty 
configuration when $\varepsilon_{d,\downarrow}(t) \sim U/2$ (this situation is however not reached for the 
largest $U$ considered). 
As this happens, of course also the local spin of the quantum dot vanishes. 
After passing through its maximum at $\Omega t= \pi/2$, $\varepsilon_{d,\sigma}(t)$ 
decreases and the quantum dot becomes again filled with a $\downarrow$ electron
when $\varepsilon_{d,\downarrow}(t) \sim U/2$.
As $\varepsilon_{d,\sigma}(t)$ continues 
to decrease towards its minimum at $\Omega t=3 \pi/2$, and additional electron  occupies the quantum dot 
when $\varepsilon_{d,\downarrow}(t) \sim -U/2$. This implies again a vanishing total spin at the quantum dot 
and as a consequence, the magnetic field does not lead to a spin polarization.  
In fact, we see in all the panels of the figure that the two occupancies differ only
for values of the gate voltage where the total mean occupancy is close to one electron, in which case, there 
is a finite spin polarization at the quantum dot, $\sum_{\sigma} n_{f, \sigma}(t) \neq 0, 2$.

We note that for $\varepsilon _{0}=\mu =0$  and any magnetic field the Hamiltonian is invariant 
under the following
transformation: $t \rightarrow -t$ and
\begin{equation}
\text{ }d_{\uparrow }^{\dagger }\rightarrow d_{\downarrow
},\text{ }d_{\downarrow }^{\dagger }\rightarrow -d_{\uparrow },  \;\;
\text{  }c_{k\uparrow }^{\dagger } \rightarrow -c_{k^\prime\downarrow },\text{ }
c_{k\downarrow }^{\dagger }\rightarrow c_{k^\prime\uparrow },    \label{transf}
\end{equation}
assuming a symmetric conduction band such that 
for any eigenstate $k$ of the isolated band, there is another one
$k^\prime$ with $\epsilon _{k^\prime} = -\epsilon _{k}$.
As a consequence of this symmetry, $n_{f,\uparrow }(t)=1-n_{f,\downarrow
}(-t)$ as can be seen in the figure.

Focusing on the interval $0 < \Omega t < \pi/2$  of Fig. \ref{nsig}, 
for which
the one-site energy of the dot increases,
we see that the expected decrease in the occupancy for the majority down spin is accompanied by 
an {\em increase} in
the occupancy of the minority up spin, denoting a charge susceptibility of opposite sign for spin up.
As already mentioned in Section IV. C, within a mean  field description, the effective local energy 
for spin up is 
$\varepsilon _{d,\uparrow}(t)=\varepsilon _{0}+\frac{\delta _{Z}}{2}+\mathcal{V}_{g}(t)
+U n_{f \downarrow}$, and the increase in $\mathcal{V}_{g}(t)$ is overcome by the decrease
in $U n_{f \downarrow}$ for large enough $U$ and $|\chi_t^{\downarrow \downarrow}|$.
For other parts of the cycle similar arguments can be followed, in particular using 
the symmetry transformation Eq. (\ref{transf}).

\begin{figure}[tbp]
\centerline{
\includegraphics[width=0.9\columnwidth]{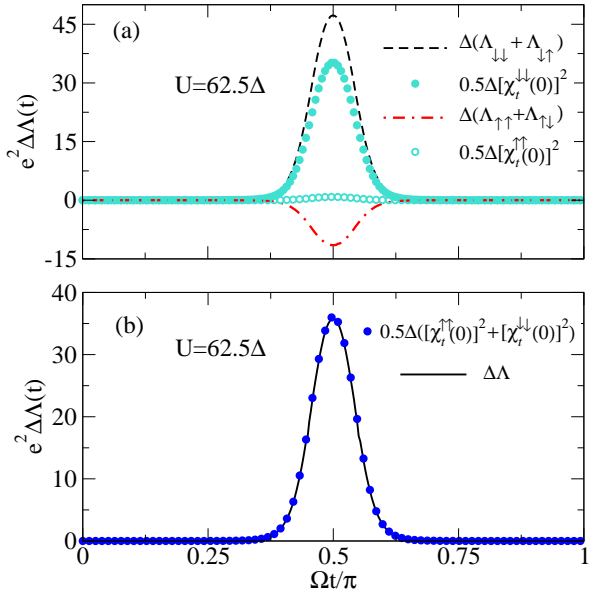}}
\caption{(Color online) Analysis of the Korringa-Shiba laws of Eq. (\ref{lamnons}) and (\ref{ks}). The 
functions $\Lambda^{\uparrow}(t)$ and $\Lambda^{\downarrow}(t)$ are  compared with
$[\chi_t^{\uparrow \uparrow}(0)]^2$ and $[\chi_t^{\downarrow \downarrow}(0)]^2$  for $U=0.05D$. Other 
parameters are the same as 
in Fig. \ref{nsig}. }
\label{gammalam}
\end{figure}

The behavior of the coefficients $\Lambda_{\sigma}$, which determine the 
non-conservative component of the  power, is illustrated 
in Fig. \ref{gammalam} for the largest value of $U$ shown 
in Fig. \ref{nsig}. These coefficients 
display a very interesting behavior as functions of time. Both exhibit features at those times where the occupancy 
of the quantum dot experiences a significant fluctuation, implying a finite charge current flowing between
the reservoir and
the quantum dot. 
The coefficient $\Lambda_{\downarrow}(t)$, associated
to electrons with the majority spin polarization is always positive
in the interval of time shown. 
Note that the symmetry transformation Eq. (\ref{transf}) implies that
$\Lambda_{\uparrow}(t)=\Lambda_{\downarrow}(-t)$.
Instead, the coefficient 
$\Lambda_{\uparrow}(t)$, which is related to the minority spin orientation can be negative.
Notice that this is in strong contrast to the non-polarized case, where the two coefficients are identical and positive.
The  coefficients $\Lambda_{\sigma}(t)$ do not separately satisfy the 
Korringa-Shiba law of the non-interacting system expressed by Eq. (\ref{lamnons}). This can be seen by 
comparing the plots in symbols with those in lines in the upper panel of the figure. However,
the total coefficient $\Lambda^c(t)$ obeys the Korringa-Shiba relation (\ref{ks}), as shown in the bottom 
panel of the figure. 

\begin{figure}[tbp]
\centerline{
\includegraphics[width=\columnwidth]{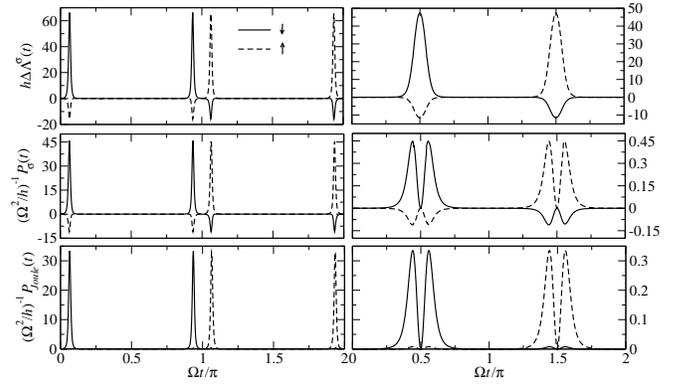}}
\caption{(Color online) Top panels: Functions $\Lambda^{\sigma}(t)$. Middle panels: Power developed by the 
forces induced by electrons with spin $\sigma$, $P_{\sigma}(t)$,
Lower panels: Joule power $P_{\rm Joule, \sigma}(t)$ (see text). Solid and dashed lines correspond to 
$\sigma= \downarrow, \uparrow$, respectively. 
Left  (right) panels correspond to 
$U=0.01D$ ($U=0.05D$), respectively. Other parameters are the same as in Fig. \ref{nsig}.  }
\label{cplmod}
\end{figure}

The corresponding behavior of the developed power is presented in Fig. \ref{cplmod} for the smallest 
and largest value of $U$ 
considered in Fig. \ref{nsig}. We show in the upper panels  the coefficients $\Lambda_{\sigma}(t)$, and 
in the
middle panels the corresponding non-conservative powers $P_{ \sigma}(t)$.
Note that $P_{ \uparrow}(t)=-P_{ \downarrow}(-t)$ as a consequence of the symmetry transformation Eq. (\ref{transf}).
For comparison, the lower panels show the IJL per spin, $P_{\rm Joule, \sigma}(t)$.
All these quantities have features at those times  $t$  where the occupancy of the dot changes 
and a charge current is established between the dot and the reservoir.  

As mentioned in the discussion of the previous figure, 
the striking feature is the different sign of $\Lambda_{\downarrow}(t)$ and $\Lambda_{\uparrow}(t)$, 
and the consequent opposite sign of the powers $P_{\downarrow}(t)$  and $P_{\uparrow}(t)$.  
This means that the contributions of different spin to the total power do not separately dissipate heat 
in the form of a Joule law, as is the case of the unpolarized quantum dot but they 
 can be decomposed as $P_{\sigma}(t)=\xi_{\sigma} P_{\rm ex}(t)+
P_{\rm Joule, \sigma}(t)$, with $\xi_{\sigma} = \pm$. Here, the component $P_{\rm Joule, \sigma}(t)$ is associated to the energy dissipated in the 
form of heat. Instead,
$P_{\rm ex}(t)$ is associated to energy that is transferred in the form of work done by the electrons with 
the minority spin component on  the electrons with the majority spin component or vice versa. The total dissipated power 
is given by
the addition of the Joule components, which is shown in the lower panels of the figure. The mechanism of 
energy exchange leading to $P_{\rm ex}(t)$ is a consequence of the combined effect of many-body interactions 
and the spin polarization
due to the magnetic field. In fact, we stress that in the non-interacting case with $U=0$,  $P_{\rm ex}(t)=0$, 
as shown in Section III B. 

To understand the fundamental difference between the non-interacting and interacting case,
let us notice that in the non-interacting case, electrons with the two spin components behave independently 
one another. Due to the Zeeman splitting there is one energy level for electrons with $\uparrow$ spin and
one for electrons with $\downarrow$ spin, which are rigidly shifted upwards and downwards in energy as the 
gate voltage changes. Every time that one of these levels gets aligned with the Fermi energy of the reservoir, 
the occupancy of the quantum dot changes and 
a current sets between the quantum dot and the reservoir. 
Such process is accompanied by Joule heating in the form of $P_{\rm Joule, \sigma}(t)$ with resistance $R_0$. 
Instead, in the interacting regime, the single occupancy is dominated by spins aligned with the magnetic field, 
while in the configurations with 0 and 2 electrons in the quantum dot, both 
spin orientations are equally populated. For this reason, when the occupancy changes from singly to double occupancy,
there is a flux of spins oriented opposed to the magnetic field following the gate voltage, 
along with a smaller counter flow of electrons aligned with the magnetic field against the gate voltage.
Analogous situations take place when the configuration changes from double to single occupancy and 
from single occupancy to the empty configuration.
The energy to generate the  current of the electrons with one of the spin components that opposes 
to the direction dictated by the external gate voltage is provided by the electrons with the opposite spin component. 
This is precisely what we have discussed within the RPA approximation
in Section IV. C. In a full cycle, this energy exchange averages to zero and only the Joule dissipation 
remains.

\begin{figure}[tbp]
\centerline{
\includegraphics[width=\columnwidth]{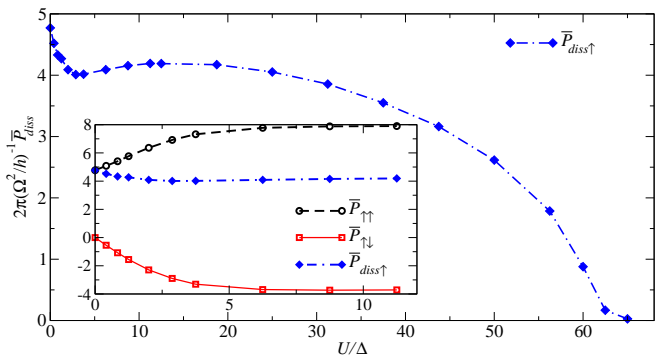}}
\caption{(Color online) Power developed by the 
forces induced by electrons with spin $\sigma$ averaged over the cycle as a function of $U$. 
The inset denotes the different components (see text) for small $U$. 
Other parameters are the same as in Fig. \ref{nsig}.  }
\label{apo}
\end{figure}

In Fig. \ref{apo} we represent the average power over the cycle for a given spin ${\bar P}_{\sigma}$.
As a consequence of the symmetry transformation Eq. (\ref{transf}) for the chosen parameters, 
${\bar P}_{\downarrow}={\bar P}_{\uparrow}$. We also represent in the figure the components
${\bar P}_{\uparrow \uparrow}$ and ${\bar P}_{\uparrow \downarrow}$, which correspond to 
the contributions of the same and opposite spin to the average total power for spin up, 
according to Eqs. (\ref{psi}), (\ref{pacad}) and (\ref{lambdass}). 
One can see that the crossed component
${\bar P}_{\uparrow \downarrow}$, which vanishes for $U=0$, decreases rapidly as $U$ is turned 
on and saturates when $U$ reaches values much larger than both $\Delta$ and the Zeeman splitting
$\delta_Z$. Instead, for small $U$, ${\bar P}_{\uparrow \uparrow}$ increases but nos so fast s the decrease
in ${\bar P}_{\uparrow \downarrow}$, so that the sum ${\bar P}_{\uparrow}$ decreases for small $U$.

For larger values of $U$ after a modest increase, ${\bar P}_{\sigma}$ decreases because the charge-transfer
peaks in the spectral density (separated by $U$) cross the Fermi level with a smaller speed, so 
that the factor $\dot{V}_g(t)^2$ is smaller [see Eq. (\ref{pdis})] and when $U$ becomes larger than
$2V_0$, the charge-transfer peaks cannot cross the Fermi energy during the cycle and the power drops to zero.

\section{Conclusions} 
We have generalized the theory of the dynamic charge
and energy transport of an interacting quantum dot modeled by an Anderson model, coupled to a reservoir 
and driven by an ac gate voltage in the
 non-linear adiabatic regime. We considered the cases with and without spin polarization due to a Zeeman 
 splitting. We have shown that the exact  adiabatic dynamics is fully characterized by the behavior of the charge susceptibility of the frozen system.
  We have presented analytical an numerical calculations obtained with NRG and 
BA techniques.

We have shown that, within and beyond linear response, and 
zero-temperature the energy is instantaneously dissipated in the form of an instantaneous Joule law
with a universal coefficient $R_0= e^2/(2 h)$ for each spin spin channel. In the case of vanishing 
magnetic field or in the absence of electron-electron interaction
there is a one-to-one correspondence between the dynamics of the driven quantum system and that of a 
classical circuit sketched in 
Fig. \ref{circuit2} with $R_0$ in each branch.
In those cases, electrons with a given spin orientation independently dissipate energy as in the circuit 
of that figure. However, in the interacting and spin-polarized case, an exchange of
energy takes place between electrons with different spin components, which cannot properly accounted by this simple classical circuit. 
This exchange of power is in some sense akin to the one between driving forces in quantum pumps discussed in 
Ref. ~ \onlinecite{exch}. 
However in that work, that mechanism takes place in non-interacting electrons driven by two 
time-dependent parameters. 
Instead, in the present case, it is a consequence of electron-electron interactions
in combination with Zeeman splitting in a system driven by a single parameter. The forces involved are non-conservative and proportional to $\dot{V}_g(t)$, like the Lorentz forces 
discussed in Ref. ~\onlinecite{lu}. However, in the present case,
the energy exchange takes place only at time intervals and do not lead to net work production when averaged over a cycle.

These predictions could be experimentally confirmed in quantum capacitors,
where, so far, only the dynamics of the charge transport has been addressed 
\cite{qcap1,qcap2,qcap3}. In fact, the Kondo regime has been realized in
similar setups without driving \cite{ford1,ford2}. Hence, the combinations
of the experimental arrays of Refs. ~\onlinecite{qcap1,qcap2,qcap3} and Refs.
\onlinecite{ford1, ford2} with suitable fast thermometry as in Refs. ~\onlinecite{therm1, therm2} should
enable the measurement of  the concomitant heat generation.

The charge susceptibility has been measured recently in a quantum dot inside an optical cavity
in the Kondo regime.\cite{compre} Our setup provides an alternative way to study this quantity.
Furthermore our works predicts non trivial results for the spin resolved components in 
the presence of strong interactions.

\section*{Acknowledgments}
J.R would like to thank Rok \v{Z}itko for his valuable
assistance in using the NRG Ljubljana code as well as E. Vernek for discussions. LA thanks M. Moskalets, D.
Sanchez and M. Filippone for useful discussions, as well as the hospitality
of the Dahlem Center for Complex Quantum Systems, FU-Berlin under the support
of the Alexander von Humboldt Foundation and the hospitality of
the ICTP-Trieste under the support of a Simons associateship. JR, PR, AAA and LA
acknowledge financial support from CONICET, MINCyT and UBACyT from
Argentina. 
AAA sponsored by PIP 112-201101-00832 of CONICET and PICT
2013-1045 of the ANPCyT.

\appendix

\section{Adiabatic Response Formalism}

For completeness we review the treatment presented in Ref. ~\onlinecite{adia}. 
Taking into account that the time-dependent perturbation ${\cal V}_g(t) n_d$ changes slowly in time, 
expanding $H(t^{\prime})$ around $t^{\prime}=t$,
the evolution operator is approximated 
up to  linear order in $\dot{\cal V}_g(t)$ as follows
\begin{equation}
\hat{U}(t,t_0)\approx T \exp\{ - i H_t(t-t_0) - i \int_{t_0}^t dt^{\prime} (t-t^{\prime}) F \dot{\cal V}_g(t)\},
\end{equation}
where $F= - \partial H(t)/\partial {\cal V}_g(t)$ is the "generalized force" induced by the driving and 
$H_t$ is the full Hamiltonian of the system frozen at time $t$.

Hence, 
to first order in $\dot{V}_{g}(t)$, using ${\cal V}_{g}(t)= e V_g(t)$,
the expectation value of an observable $A$ can be expressed in
this ``adiabatic approximation'' as: 
\begin{equation}
\langle A(t)\rangle \approx \langle A\rangle_{t} -\frac{i e}{\hbar}
 \int_{t_{0}}^{t}\left(t-t^{\prime }\right)\langle[A(t),F(t^{\prime })]
\rangle_t dt^{\prime }\dot{V}_{g}(t),  \label{AdiabApprox}
\end{equation}
with $\langle A\rangle_{t}=\text{Tr}[\rho_t A]$, and the operators $F(t),\;A(t)$ 
evolving according to the Heisenberg picture with respect to 
$H_t$, i.e., $o(t)=e^{\frac{iH_t t}{\hbar}}oe^{-\frac{iH_t t}{\hbar}}$. From
Eq.~(\ref{AdiabApprox}), 
and using $F(t)=-n_d(t)$,
an adiabatic retarded susceptibility corresponding
to the frozen Hamiltonian $H_t$ can be defined as $\chi^A_{ t}(t-t^{\prime})=- i \theta(t-t^{\prime })\langle[A(t),n_d(t^{\prime})]\rangle_t$. Hence, 
\begin{equation}
\langle A(t)\rangle \approx \langle A\rangle_{t} + e \Lambda_{t} \dot{V}_{g}(t),
\end{equation}
with 
\begin{equation}  \label{lambda2}
\Lambda_{t}=-\frac{1}{\hbar} \int_{t_{0}}^{\infty}\left(t-t^{\prime }\right)
\chi^A_{ t}(t-t^{\prime }) dt^{\prime }.
\end{equation}
In Eq.~(\ref{lambda2}) let us make the change of variables $\tau=t-t^{\prime }$. 
We have, taking $t_0= -\infty$,
\begin{equation}  \label{lambdathree}
\Lambda_{t}= -\frac{1}{\hbar} \int_{-\infty}^{\infty}\tau \chi^A_{ t}(\tau)
d \tau.
\end{equation}
We define the Fourier transform with respect to $\tau= t- t^{\prime}$ 
\begin{equation}
\chi^{A}_t(\omega)= \int d\tau e^{i \omega \tau} \chi_t^A(\tau)
\end{equation}
We can now use 
\begin{equation}
\frac{\partial \chi^{A}_t(\omega)}{\partial \omega}=i \int d\tau e^{i \omega
\tau} \tau \chi_t^A(\tau)
\end{equation}
along with the fact that $\Lambda_t$ is a real function to write 
\begin{equation}
\Lambda_t= - \frac{1}{\hbar} \left.\frac{d~ \mathrm{Im}[\chi^{A}_t(\omega)] }{
d\omega}\right|_{\omega=0} = - \frac{1}{\hbar} \lim_{\omega \rightarrow 0} 
\frac{ \mathrm{Im}[\chi^{A}_t(\omega)] }{\omega},
\end{equation}

where in the last step we have assumed that $\mathrm{Im}[\chi_{A}(0]=0$.
Notice that $\omega$ has units of frequency, thus in order to compare with
NRG results one has to transform to energy units by making $\omega=\epsilon/\hbar$.

\section{Equation for the non-linear circuit} \label{eqcirc}
Here we analyze the current through each branch of the classical circuit
of Fig. \ref{circuit2}.
We define the differencial capacitance from 
\begin{equation}
C_{\sigma }(V_{C}^{\sigma })=\frac{dq_{\sigma }}{dV_{C}^{\sigma }},
\label{cap}
\end{equation}%
where $q_{\sigma }$ is the charge in the capacitor and $V_{C}^{\sigma }$ the
potential drop through it. The total potential drop is $V=V_{C}^{\sigma
}+V_{R}^{\sigma }$, where $V_{R}^{\sigma }=R_{\sigma }I_{\sigma }$ is the
potential drop through  the resistance and $I_{\sigma }$ is the current in
the branch with spin $\sigma $. 
Using $I_{\sigma }=dq_{\sigma }/dt=\dot{q}_{\sigma }$ and Eq. (\ref{cap}) 
one has a differential equation for the
current,

\begin{equation}
I_{\sigma }=C_{\sigma }(V-R_{\sigma }I_{\sigma })\left[ \dot{V}-\frac{%
d(R_{\sigma }I_{\sigma })}{dt}\right] .  \label{de}
\end{equation}

The derivatives with respect to the time introduce a factor of the order of
the frequency $\Omega $. We now assume small frequencies, specifically $%
\Omega R_{\sigma }I_{\sigma }\ll 1$. Then as a first approximation, the last
term of Eq. (\ref{de}) can be neglected. This limit also implies $V\gg
R_{\sigma }I_{\sigma }$, an thus to the same order of approximation one can
evaluate take $C_{\sigma }(V)$ as the first factor of  Eq. (\ref{de})
leading to

\begin{equation}
I_{\sigma }^{1}=C_{\sigma }(V)\dot{V},  \label{i1}
\end{equation}%
to first order in $\dot{V}$. Replacing this result in the second member of
Eq. (\ref{de}), expanding the first factor and using 
$dC_{\sigma }/dt=\dot{V}dC_{\sigma }/dV$.
we obtain the following expression for the current to second order in $\dot{V},$

\begin{equation}
I_{\sigma }=C_{\sigma }\dot{V}-\frac{d(R_{\sigma }C_{\sigma }^{2}\dot{V})}{dt}.  
\label{circuit1}
\end{equation}%

In order to map the classical circuit to our quantum problem, we identify 
$V= V_{g}$ as minus the gate voltage of the quantum dot (the energy of electrons with negative 
charge in the dot changes as the gate voltage increases) and  according to the definition Eq. (\ref{ic}) 
$I_\sigma=dq_{\sigma }/dt=-I_{C,\sigma}$.
Performing the corresponding replacements one obtains  Eq. (\ref{circuit}).

\section{Numerics}

In this work we use the known ``NRG Ljubljana'' free code to calculate the
observables of interest (and their correlations) in the case where there is
a local Coulomb interaction in the dot ($U>0$) and also for $U=0$. There are
some subtleties in the NRG method, most of them described in
the literature (see for instance Ref. \onlinecite{Bulla2008} and
references therein). However, for the sake of completeness, the main steps
of the NRG approach we have employed are described below.

\

1) We first map the Anderson model onto a Wilson chain, composed of an
impurity site with its many body term plus a linear chain of non-interacting
sites. Upon this process, the continuous Anderson model has been discretized
in the energy space. Here we used the known $z$-trick logarithmic
discretization scheme, in which the characteristic energy scale is given by%
\cite{Zitko2009,Campo2005,Yoshida1990}
\begin{eqnarray}
\epsilon_N=\frac{1-\lambda^{-1}}{\log \lambda}\lambda^{-(N-1)/2+1-z},
\end{eqnarray}
where here we have set $\lambda=2$ and averaged the results for 32 values of
$z \in [0,1]$.

\

2) We then diagonalize iteratively a series of Hamiltonians, starting from
an initial $H_{-1}$ describing just the impurity (containing the many-body
term) and increasing it with a non-interacting site of the Wilson chain at
each iteration. Hence, at a given iteration we need to diagonalize a
Hamiltonian $H_N$ describing the impurity plus $N+1$ sites of the chain
whose dimension is $(N+2)\times (N+2)$. Because of the rapid increase of the
dimension of the Hilbert space along the iterating process, we have to take
advantage of the symmetries of the Hamiltonians. Here, Since $H_{N}$
commutes with both the total charge ($\hat Q_N$) of chain described by $H_N$
and with its total spin operator squared ($\hat S_N^2$), we exploit U(1)$%
\times$ SU(2) symmetry. In this way, at the $N$-th NRG iteration we
diagonalize an enlarged block-diagonal $(N+2)\times (N+2)$ Hamiltonian
matrix 
whose sectors are labeled by the quantum number ($Q_N,S_N$), where $Q_N$ and
$S_N$ represents the eigenvalues of $\hat Q_N$ and $\hat S_N^2$,
respectively.

Even though taking advantage of the U(1)$\times$ SU(2) is a
great improvement, it is not enough to allow us to diagonalize large Wilson
chains. To overcome this problem it is necessary to truncate the Hilbert
space by discarding states. %
resolution of
%
In the present calculations we retain states with energy up to $E_{\text{keep%
}}=2\epsilon_N$, a reasonably good choice to converge the many-particle
eigen-energies to the strong coupling fixed point.
i.e., it does not change upon
reached the fixed point of the
At each iteration we not
only diagonalize the Hamiltonians but also calculate all the physical
quantities we are interested in.

\

3)
With the matrix elements of the relevant quantities at each iteration,
we can calculate the thermodynamic and 
dynamic quantities, such as the static and dynamic
susceptibilities. For the dynamic quantities we have employed the full
density matrix (FDM) version of the NRG, which is known to provide a
better resolution of the spectral quantities.
The energy delta peaks
appearing in the dynamic susceptibilities are usually broadened by
using various
smooth distribution functions \cite{Bulla2008}. In our case we use a
modified broadening kernel $K(\epsilon,\epsilon_{j})$
defined piecewise by \cite{Osolin2013}

\begin{equation}
K(\epsilon,\epsilon_{j})=
\begin{cases}
L(\epsilon,\epsilon_j) & \text{if } \vert \epsilon_j\vert\geq \epsilon_0 ,
\\
G(\epsilon,\epsilon_j)\left[1-h(\epsilon)\right] & \text{if }\vert
\epsilon_j\vert \leq \epsilon_0,%
\end{cases}%
\end{equation}
with

\begin{align}
G(\epsilon,\epsilon_{j})&=\frac{\theta(\epsilon \epsilon_j)} {\sqrt{\pi}%
\alpha \vert\epsilon\vert}\text{exp}\left[-\ln\left(\frac{\vert
\epsilon/\epsilon_j\vert}{\alpha}-\gamma\right)^2\right], \\
L(\epsilon,\epsilon_{j})&=\frac{1}{\sqrt{\pi} \epsilon_0}\text{exp}\left[
-\ln\left(\frac{\vert \epsilon/\epsilon_0\vert}{\alpha}\right)^2\right], \\
h(\epsilon)&=\text{exp}\left[-\ln\left(\frac{\vert \epsilon/\epsilon_0\vert}{%
\alpha} \right)^2\right],
\end{align}

where $\alpha$ defines the broadening parameter, $\gamma=\alpha/4$,
and $\omega_0$
is an energy threshold that changes the broadening
distribution function from a Log-Gaussian to a Gaussian at low energies.
In practice, smaller  $\alpha$ diminishes NRG
over broadening, but leads to non-physical
oscillations in the dynamic susceptibilities, which can be reduced by
averaging over a convenient number of discretization meshes of the
conduction band. The broadening parameter is chosen to be $\alpha=0.02$ and $%
\epsilon_0=10^{-99}D$.




\begin{thebibliography}{99}
\bibitem{qcap1} J. Gabelli, G. F\`eve, J.-M. Berroir, B. Placais, A.Cavanna,
E.al,Y. Jin, and D. C. Glattli, Science \textbf{313}, 499 (2006).

\bibitem{qcap2} G. F\`eve, A. Mah\'e, J.-M. Berroir, T. Kontos, B. Placais,
C. Glattli, A. Cavanna, B. Etienne, and Y. Jin, Science \textbf{316}, 1169
(2007).

\bibitem{qcap3} J. Gabelli, G. F\'eve, J.-M. Berroir, and B. Placais, Rep.
Prog. Phys. \textbf{75}, 126504 (2012).

\bibitem{btp1} M. B\"uttiker, A. Pr\^etre, and H. Thomas, Phys. Rev. Lett. 
\textbf{70}, 4114 (1993).

\bibitem{btp2} A. Pr\^etre, H. Thomas, and M. B\"uttiker, Phys. Rev. B 
\textbf{54}, 8130 (1996).

\bibitem{btp3} M. B\"uttiker, H. Thomas, and A. Pr\^etre, Phys. Lett. A 
\textbf{180}, 364 (1993).

\bibitem{rosa1} S. E. Nigg, R. L\'opez, and M. B\"uttiker, Phys. Rev. Lett. 
\textbf{97}, 206804 (2006).

\bibitem{rosa2} M. Lee, R. L\'opez, M.-S. Choi, T. Jonckheere, and T.
Martin, Phys. Rev. B \textbf{83}, 201304 (2011).

\bibitem{mora} C. Mora and K. Le Hur, Nat. Phys. \textbf{6}, 697 (2010).

\bibitem{miche1} M. Filippone, K. Le Hur, and C. Mora, Phys. Rev. Lett. 
\textbf{107}, 176601 (2011).

\bibitem{miche2} M. Filippone, C. Mora, Phys. Rev. B \textbf{86}, 125311
(2012).

\bibitem{nonlin} M. Moskalets, P. Samuelsson, and M. B\"uttiker Phys. Rev.
Lett. \textbf{100}, 086601 (2008).

\bibitem{misha1} S. Okhovskaya, J. Splettstoesser, M. Moskalets, and M.
B\"uttiker, Phys. Rev. Lett. \textbf{101}, 166802 (2008).

\bibitem{misha2} J. Splettstoesser, M. Moskalets, and M. B\"uttiker, Phys.
Rev. Lett. \textbf{103}, 076804 (2009).

\bibitem{misha3} G\'eraldine Haack, Michael Moskalets, Markus B\"uttiker
Phys. Rev. B \textbf{87}, 201302(R) (2013).

\bibitem{janine} O. Kashuba, H. Schoeller, J. Splettstoesser, Eur. Phys.
Lett. \textbf{98}, 57003 (2012).

\bibitem{david} M. I. Alomar, J. S. Lim, and David S\' anchez, Phys. Rev. B {\bf 94}, 165425 (2016).

\bibitem{lbf}D. Litinski, P. W. Boruwer, and M. Filippone, arXiv:1612.04822.

\bibitem{Ludovico2014} M. F. Ludovico, J. S. Lim, M. Moskalets, L. Arrachea
and D. S\' anchez, Phys. Rev. B \textbf{89}, 161306(\textbf{R}) (2014).

\bibitem{exch}L. Arrachea, M. Moskalets and L. Martin-Moreno, Phys. Rev. B {\bf 75} 245420 (2007).

\bibitem{ludo16} M. F. Ludovico, M. Moskalets, D. S\' anchez, and L.
Arrachea, Phys. Rev. B \textbf{94}, 035436 (2016).

\bibitem{note} Notice that this definition is consistent with
$\dot{U}= \dot{Q}- P$, where $U$ is the internal energy of the electron system (assumed to remain constant for the full system containing the dot and the reservoir).


\bibitem{kondo} A. C. Hewson, "The Kondo Problem to Heavy Fermions"
(Cambridge University Press, Cambridge, 1993).

\bibitem{adia} M.F. Ludovico, F. Battista, F. von Oppen and L. Arrachea,
Phys. Rev. B \textbf{93}, 075136 (2016).

\bibitem{lu} J-T. L\"u, M. Brandbyge and P. Hedegard, Nano Lett. {\bf 10} 1657 (2010).

\bibitem{korr-shiba} H. Shiba, Prog. of Theor. Phys. \textbf{54}, 967 (1975).

\bibitem{lili} L. Arrachea, Phys. Rev. B \textbf{72}, 125349 (2005); ibid 
\textbf{75}, 035319 (2007).

\bibitem{wt} P. B Wiegmann and A. M. Tsvelick, J. Phys. C \textbf{16}, 2281 (1983).

\bibitem{anda} I. J. Hamad, P. Roura-Bas, A. A. Aligia, and E. V. Anda,  
Physica Status Solidi (b) \textbf{253}, 478 (2015).


\bibitem{Zitko2009} R. \v{Z}itko, NRG LJUBLJANA , 
open source numerical renormalization group code, 
http://nrgljubljana.ijs.si

\bibitem{arf} A. A. Aligia, P. Roura-Bas, and S. Florens, Phys. Rev. B \textbf{92}, 035404 (2015); 
references therein.

\bibitem{compre}M. M. Desjardins, J. J. Viennot, M. C. Dartiailh, L. E. Bruhat, M. R. Delbecq, M. Lee, M.-S. Choi, A. Cottet, and  T. Kontos, Nature {\bf 545}, 71 (2017).

\bibitem{Campo2005} V.L. Campo and L.N. Oliveira, Phys. Rev. B \textbf{72},
104432 (2005).

\bibitem{Yoshida1990} M. Yoshida, M. A. Whitaker, and L. N. Oliveira, Phys.
Rev. B \textbf{41}, 9403 (1990).

\bibitem{Bulla2008} R. Bulla, T. A. Costi, and T. Pruschke, Rev. Mod. Phys. 
\textbf{80}, 395 (2008).

\bibitem{Osolin2013} Z. Osolin, and R. \v{Z}itko, Physical Review B \textbf{87},
245135 (2013).

\bibitem{ford1} M. Kataoka, C. J. B. Ford, M. Y. Simmons, and D. A. Ritchie
Phys. Rev. Lett. \textbf{89}, 226803 (2002).

\bibitem{ford2} H.-S. Sim, M. Kataoka, C.J.B. Ford, Phys. Rep. \textbf{456},
127 (2008).

\bibitem{therm1} S. Gasparinetti, K. L. Viisanen, O.-P. Saira, T. Faivre, M.
Arzeo, M. Meschke, and J. P. Pekola Phys. Rev. Applied \textbf{3}, 014007
(2015).

\bibitem{therm2} 
    M. Zgirski, M. Foltyn, A. Savin, M. Meschke, J. Pekola, arXiv:1704.04762.

\end{thebibliography}
\end{document}